\documentclass{article}
\usepackage{picins}
\usepackage{spconf,amsmath,graphicx}
\usepackage[table,xcdraw]{xcolor}
\usepackage{utfsym}
\usepackage{hyperref}

\hypersetup{
    colorlinks=true,
    linkcolor=blue,
    filecolor=magenta,      
    urlcolor=blue,
    pdfpagemode=FullScreen,
}
\usepackage{enumitem}
\setlist{nosep, leftmargin=14pt}

\usepackage{mwe} 
\usepackage{multirow}
\usepackage{graphicx}
\usepackage{booktabs}

\title{APIS: A paired CT-MRI dataset for ischemic stroke segmentation challenge}

\name{%
\begin{tabular}{@{}c@{}}
    Santiago Gómez$^{\star}$, 
    Daniel Mantilla$^{\dagger}$, 
    Gustavo Garzón$^{\star}$,\\
    Edgar Rangel$^{\star}$, 
    Andrés Ortiz$^{\dagger}$, 
    Franklin Sierra$^{\star}$, 
    Fabio Martínez$^{\star}$
\end{tabular}}

\address{Biomedical Imaging, Vision, and Learning Laboratory (BIVL$^2$ab) -\\ Universidad Industrial de Santander (UIS), Colombia$^\star$ \\ Clínica FOSCAL, Floridablanca, Colombia$^\dagger$}

\begin{document}

\maketitle

\begin{abstract}

Stroke is the second leading cause of mortality worldwide. Immediate attention and diagnosis play a crucial role regarding patient prognosis. The key to diagnosis consists in localizing and delineating brain lesions. Standard stroke examination protocols include the initial evaluation from a non-contrast CT scan to discriminate between hemorrhage and ischemia. However, non-contrast CTs may lack sensitivity in detecting subtle ischemic changes in the acute phase. As a result, complementary diffusion-weighted MRI studies are captured to provide valuable insights, allowing to recover and quantify stroke lesions. This work introduced APIS, the first paired public dataset with NCCT and ADC studies of acute ischemic stroke patients. APIS was presented as a challenge at the 20th IEEE International Symposium on Biomedical Imaging 2023, where researchers were invited to propose new computational strategies that leverage paired data and deal with lesion segmentation over CT sequences. Despite all the teams employing specialized deep learning tools, the results suggest that the ischemic stroke segmentation task from NCCT remains challenging. The annotated dataset remains accessible to the public upon registration, inviting the scientific community to deal with stroke characterization from NCCT but guided with paired DWI information.
\end{abstract}



\section{Introduction}
Stroke is the second leading cause of mortality worldwide and the most significant adult disability in developed countries \cite{feigin2022world}. A dramatic projection estimates that one in four people over 25 years will suffer a stroke during their lifetime \cite{roth2020global}. Ischemic stroke, caused by blood vessel occlusion, is the most prevalent type of stroke (80\% of all cases) with a significant risk of morbidity \cite{rennert_epidemiology_2019}. A dramatic point is that brain tissue is extremely sensitive to ischemia, producing irreversible damage within minutes from the onset. Therefore, targeted therapies must be delivered within minutes to hours from symptom onset for maximum effectiveness \cite{powers2019guidelines}.

Typically, the first-line imaging modality in most centers is a non-contrast CT (NCCT), allowing to exclude hemorrhagic stroke and intracranial hemorrhage. The NCCT is a low-cost, fast, and accessible modality, bringing adequate conditions for early analysis. Nonetheless, this modality reports low sensitivity with respect to ischemia in the early stages of the disease. To overcome such NCCT limitations, clinical protocols include diffusion-weighted imaging (DWI) sequences, such as DWI B-1000 and the apparent diffusion coefficient (ADC), which are biomarkers of cytotoxic injury that predict edema formation and outcome after ischemic stroke \cite{bevers2018apparent}. Hence, the standard lesion analysis requires neuroradiologists to observe both modalities to derive a comprehensive lesion characterization. However, this observational task is challenging, time-consuming (taking approximately 15 minutes per case), and susceptible to bias errors \cite{Martel1999,rana2003apparent}.

Currently, there is a need to formulate computational approaches that allow physicians to carry out the analysis of stroke lesions, reducing the bias from expert observations over NCCT, allowing rapid decisions on the appropriateness of interventional treatments (i.e., mechanical thrombectomy or thrombolysis) for stroke patients.

Public datasets for the segmentation of ischemic stroke from different image modalities have been released since 2015 \cite{maier_isles_2017,winzeck2018isles,hakim2021predicting,petzsche2022isles,liew2018large,liew2022large,de_bruijne_symmetry-enhanced_2021}. Thanks to the availability of such public datasets, the literature has significantly increased in the number of research proposals to support ischemic stroke lesion segmentation. More specifically, several works have leveraged public datasets with parametric maps computed from conventional thresholding analyses on CT perfusion (CTP) \cite{hakim2021predicting}. Existing approaches have used multi-context representations, encoded from deep U-net-like architectures, to generate segmentation masks \cite{Tureckova2019,Kuang2021}. In the first version of such dataset were included DWI sequences (taken as inputs for delineations).  Such modality allowed a better lesion segmentation from enriched deep representations, following synthetic approximations \cite{Wang2020}. Nonetheless, DWI images are no longer available, reducing the possibility of exploiting new synthetic strategies. Isolated efforts over such translation schemes have been reported between FLAIR and NCCT images, but following owner datasets \cite{gutierrez_lesion-preserving_2023}. 

From an alternative public dataset with only NCCT studies, some computational approaches modelled the anatomical symmetry to compute differences between hemispheres and estimate ischemic stroke lesions from pathological asymmetries \cite{de_bruijne_symmetry-enhanced_2021,kuang2019automated,wang2016deep}. Alternative approaches have characterized stroke from datasets with isolated MRI studies, motivated by the high sensitivity and specificity of DWI sequences \cite{liew2018large,winzeck2018isles,petzsche2022isles}. Despite multiple efforts in the community to provide stroke studies from CT sequences, the characterization of lesions remains as an open problem due to the complexity of low-contrast observations and the lack of CT paired with other modalities to guide segmentation during training \cite{dekeyzer2017distinction}.

This work presents \textit{APIS: A Paired CT-MRI dataset for Ischemic Stroke Segmentation}, the first publicly available dataset featuring paired CT-MRI scans of acute ischemic stroke patients, along with lesion annotations from two expert radiologists. This dataset was introduced as a challenge at the 20th IEEE International Symposium on Biomedical Imaging (ISBI) in 2023. A total of 60 paired studies containing NCCT and ADC image modalities and delineations from an expert radiologist, were released for training. As for the testing stage, we included 36 studies comprising NCCT scans and delineations from two expert radiologists. The paired modalities allow the exploration of complementary findings proper of each modality and its impact on the stroke lesion segmentation task. Additionally, a further agreement ana\-ly\-sis could be performed with the expert's annotations in order to estimate the model generalization capability, avoiding the bias expert's annotations. In this work is also reported a validation and analysis of the achieved results using different segmentation strategies in the context of the APIS challenge.

\begin{figure*}[ht!]
    \centering
    \includegraphics[width=\textwidth]{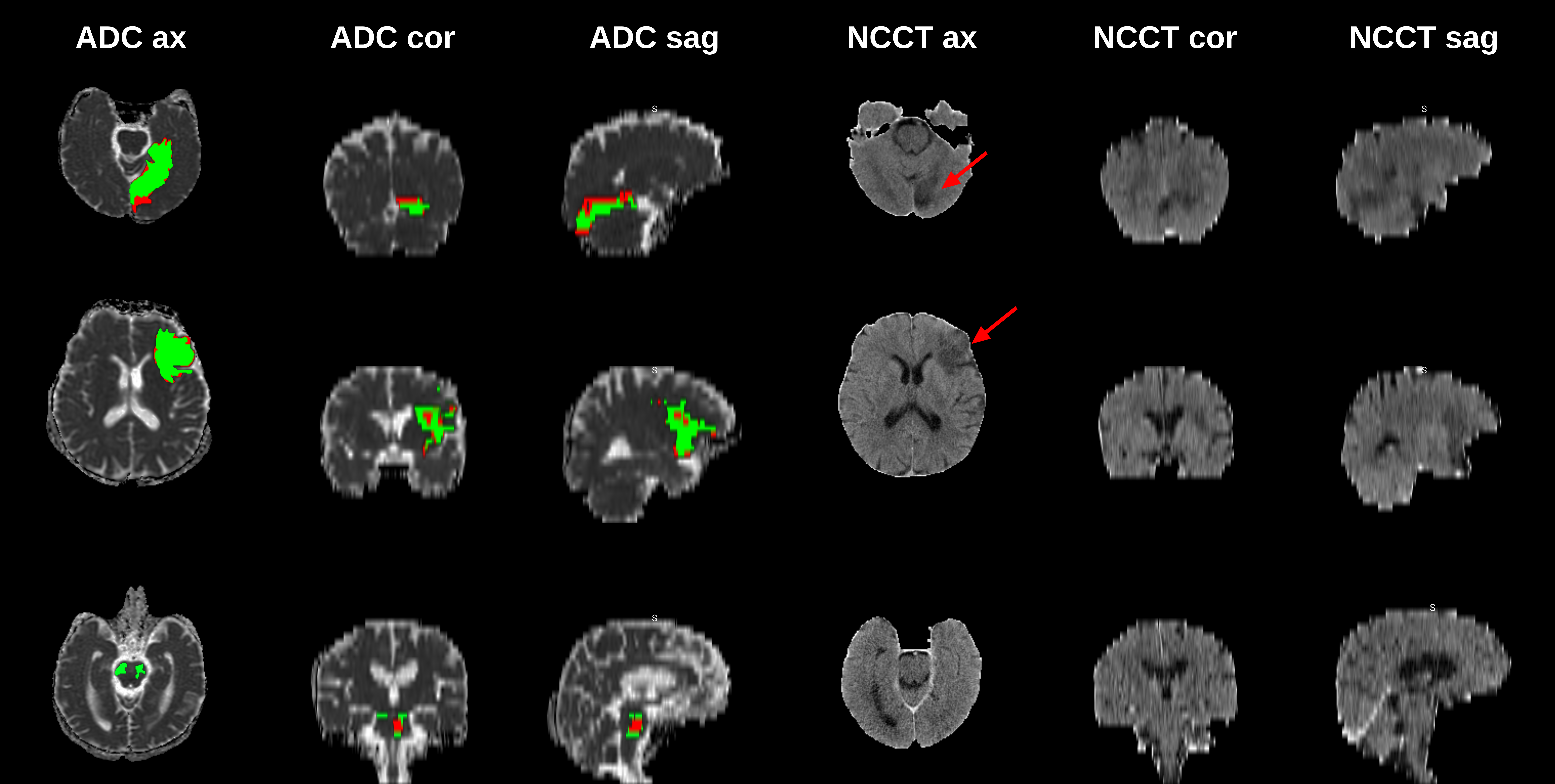}
    \caption{Examples of image modalities and annotations in the APIS dataset. In the three leftmost columns are shown ADCs in axial, coronal and sagittal planes, with the annotations of two experts. Complementary, the same views are shown for the NCCT modality in the three rightmost columns. In the top two columns can be observed lesions that appear hypointense on ADC and hypoattenuated on NCCT. Contrarily, the last row shows an example where the lesion is hard to see on NCCT.}
    \label{information_dataset}
\end{figure*}

\section{Acute ischemic stroke segmentation from CT}

Thanks to the availability of public data related to stroke image studies, several computational strategies have been proposed to reduce subjectivity in shape lesion estimation. These strategies have the capability to learn data's variability to support the segmentation and characterization of affected brain regions. In this line, the AISD dataset \cite{de_bruijne_symmetry-enhanced_2021} provides around 397 NCCT studies, which were delineated by one expert radiologist. From this dataset, some approaches have integrated anatomical symmetry information, for instance, exploiting the pixel-wise difference between hemispheres \cite{kuang2019automated,wang2016deep}. Also, from such dataset have been proposed unsupervised nets to obtain bilateral aligned images, standing out in-planar and across-planar symmetry\cite{de_bruijne_symmetry-enhanced_2021}. Likewise, Ni \textit{et al.} \cite{wang_asymmetry_2022} proposed three 3D CNNs to sequentially align NCCT inputs across the mid-sagittal axis, generate an image with more salient pathological asymmetries, and to segment the ischemic stroke lesions. Contrarily, Xu \textit{et al.} \cite{xu_combining_2023} obtained competitive results without explicitly modeling anatomical symmetry information. Specifically, they implemented a residual convolutional autoencoder with six deformable transformer layers in the bottleneck to segment ischemic lesions from 3D NCCT inputs. These approaches learn complex radiological findings with the capability to discriminate between lesion and healthy tissue. However, such discrimination remains limited due to the observed low signal-to-noise and contrast-to-noise ratios of brain tissues, preventing computational nets from learning more sophisticated descriptors. 

Moreover, the ISLES2018 dataset \cite{hakim2021predicting} provides 156 CTP studies from acute ischemic stroke patients, of which 64 were kept for a hidden test. The infarcts were delineated from DWI images, taken within 3 hours from CTP. Perfusion studies allow to quantify blood flow through the brain parenchyma, potentially indicating brain damage due to stroke. From such studies are computed parametric maps that summarize the dense information contained in raw CTP. Particularly, the parametric maps included are cerebral blood flow (CBF), cerebral blood volume (CBV), mean transit time (MTT), and time to maximum (TMax), calculated by conventional thresholding methods. Regarding ISLES2018, some studies have used perfusion-derived parametric maps to obtain an improvement in the segmentation of the ischemic stroke lesions. For instance, Tureckova \textit{et al.} used a 3D U-Net based model with dilated convolutions \cite{Tureckova2019}, while Dolz \textit{et al.} proposed an U-Net architecture that take image modalities separately with multiple dilated convolutional encoders \cite{Dolz2019}. These representations remain limited to recover whole context information due to error propagation on skip connections. Furthermore, Clèrigues \textit{et al.} proposed a 2D asymmetric residual encoder–decoder framework, following a dynamically weighted loss function and mini-batches of image patches \cite{Clerigues2019} to deal with the natural class imbalance of stroke segmentation. Other approaches have integrated MRI-DWI sequences to enrich observations taken from CTP studies. For instance, Liu \textit{et al.} \cite{Liu2019} proposed an adversarial-generation of synthetic MRI-DWI images from CTP studies. The synthetic DWI image is then mapped through a  segmentation network to predict the lesion outcome. Similarly, Wang \textit{et al.} \cite{Wang2020} proposed a three network pipeline to generate synthetic DWI sequences from 4D Perfusion CT sequences. Despite their remarkable performance, these studies cannot be reproduced because paired and aligned DWIs are no longer available. Moreover, obtaining such paired CT-MRI studies within a short-time window is challenging in clinical scenarios. The presence of CTP maps allowed computational approaches to improve their lesion segmentation capabilibities. Nevertheless, obtaining valid perfusion maps requires to obtain appropriate arterial input functions (AIFs), and venous output functions (VOFs) can affect both visual and quantitative assessments of perfusion CT metrics \cite{allmendinger2012imaging}.

Other computational alternatives have used private da\-ta\-bases with relevant clinical variables and scales, captured during the attention protocol. For instance, Kuang \textit{et al.} proposed a multi-task learning strategy to simultaneously segment early infarcts and score the Alberta Stroke Program Early CT Score (ASPECTS) from an owner dataset with 260 NCCT studies \cite{Kuang2021}. The strategy consist of encoding the brain symmetry from NCCT inputs information using three convolutional encoders. Next, a multi-level attention gate module (MAGM) is used to decode the early infarct correlated activations for the segmentation and ASPECTS networks. Furthermore, Gauriau \textit{et al.} implemented a classical UNet to segment acute ischemic lesions from 3556 NCCTs studies \cite{gauriau_head_2023}. Noteworthy, they reported that their model outperformed experts with more than 25 years of experience. Such strategies, although novel, are isolated proposals that leverage private data, imposing difficulties to reproduce and validate them. Moreover, these works missed out the opportunity to exploit paired CT-MRI information to improve the ischemic stroke lesion segmentation capabilities of computational approaches.

\section{The APIS dataset and challenge}
This work introduces a paired CT-MRI dataset, carefully built to exploit complementary radiological findings and support stroke lesion segmentation. This dataset was presented in the ISBI official challenge \textit{"APIS: A Paired CT-MRI Dataset for Ischemic Stroke Segmentation Challenge"}\footnote[1]{Official challenge webpage: \href{https://bivl2ab.uis.edu.co/challenges/apis}{https://bivl2ab.uis.edu.co/challenges/apis}}. To build the dataset, a retrospective study was conducted to collect 96 studies of patients presenting with stroke symptoms at two clinical centers between October 2021 and September 2022. After reviewing imaging studies, the studies were categorized into control (n=10) or ischemic stroke (n=86) studies. Control patients with stroke symptoms were included to diversify tissue samples, potentially enhancing deep learning models' ability to segment stroke lesion tissue. For each study, the triage NCCT and the subsequent ADC were captured. The inclusion criteria involve patients older than 18 years, without signs of cerebral hemorrhage, with no treatment between NCCT and MRI image acquisition, and confirmed ischemic stroke diagnosis after reviewing imaging studies. Also, if there was evidence of partial reperfusion, patients were not excluded. 

Table \ref{current_data} shows the demographic data used for the train and test partitions. The studies were captured using two CT scanners: i) a Toshiba Aquilion 64 TSX-101 and ii) a Toshiba Aquilion one TSX-301/ac scanner, with 64 and 320 detectors, respectively. Also, two MRI scanners, namely i) Toshiba Vantage Titan MRI and ii) General Electric SIGNA Explorer. The ischemic lesion annotations were carried out by two neuro-interventional radiologists with more than five years of experience. Firstly, experts reviewed clinical variables. Subsequently, they identified if it was possible to identify a lesion from the NCCT. Finally, experts assessed DWI and ADC images and carried out the manual delineations on ADC. The lesion annotation process was performed by each radiologist individually using the MRICroGL software \cite{li2016first}. Figure \ref{information_dataset} shows some examples of APIS dataset, together with expert delineations.

\begin{table}[h!]
\centering
\scalebox{0.64}{
    \begin{tabular}{ccc}
    \toprule
    \textbf{VARIABLE}                  & \textbf{TRAIN}             & \textbf{TEST}               \\ 
    \midrule
    \textbf{Age {[}Range{]}(mean $\pm$ std)}     & {[}30, 94{]}(70.23 $\pm$ 12.81) & {[}22, 93{]}(73.06 $\pm$ 15.46) \\
    \textbf{Genre (Female / Male)}           & 28 / 32                    & 19 / 17                     \\
    \textbf{Hypertension (Yes / No)}         & 37 / 23                    & 21 / 15                     \\
    \textbf{Smoking (Yes / No)}              & 3 / 57                     & 35 / 1                      \\
    \textbf{Diabetes (Yes / No / Unk)}             & 15 / 45 / 0                    & 14 / 21 / 1                 \\
    \textbf{Previous Stoke(Yes / No / Unk)}  & 9 / 49 / 2                 & 4 / 31 / 1                  \\
    \textbf{NIHSS  {[}Range{]}(mean $\pm$ std)}  & {[}0, 29{]} (6.68 $\pm$ 6.69)  & {[}0, 22{]} (8.0 $\pm$ 6.28)    \\
    \textbf{Laterality (Left / Right / Unk)} & 33 / 25 / 2                & 21 / 15 / 0                  \\ 
    \textbf{Institution (FOSCAL / FOSUNAB)}  & 54 / 6                     & 35 / 1                  \\ 
    \bottomrule
    \end{tabular}
}
\caption{Demographic data for the train and test splits in the APIS challenge dataset.}
\label{current_data}
\end{table}

All studies were skull-stripped and co-registered w.r.t. labeled ADC sequences. Figure \ref{information_dataset} shows a preview of the image modalities and the respective segmentation mask. The delineated lesions show high variability in shape, size, and number of consecutive slices where the lesion occurs.

Regarding the challenge, the participants were invited to propose computational approaches that take advantage of paired CT-ADC studies to codify patterns observed from CT. Then, during evaluation, the computational proposal is measured regarding the capability to segment lesions over CT. For training, the participant counts with a total of 60 CT-MRI studies while for the test set, the participants have 36 CT studies. To measure the performance of the submitted algorithms, overlap metrics such as the dice score (DICE), precision (PREC), and sensitivity (SENS) were used to measure the overlap between a reference mask and the algorithm prediction. Furthermore, the successful detection of a lesion and its boundaries were assessed. The Hausdorff Distance (HD), with an Euclidean distance measure, was used to assess the accuracy of lesion boundaries. Participants were invited to submit their solution, in order to predict the stroke lesions over an unseen hidden test. Algorithm predictions were compared to two manual expert delineations.

\subsection{Ethics statement}
The local ethics boards of all participating centers granted approval for the retrospective assessment of imaging data. To ensure complete anonymity, all patient information was eliminated from the volumetric nifty files, and any facial features were removed through skull stripping. Due to the retrospective nature of the study and the de-identification of patient data, the requirement for written informed consent was exempted by the ethics boards.


\section{Evaluation and results}


The APIS challenge was open in December 2022 and received inscriptions for a total of 41 teams. Training data was released in December 26th of 2022, and during four months the teams had the opportunity to design and implement strategies to segment acute ischemic stroke lesions from the NCCT images. For test, a total of four teams submitted a solution. Finalist teams were: \textit{icomaia}\footnote[2]{Universitat de Girona and icometrix}, \textit{xmu\_lsgroup}\footnote[3]{Xiamen University}, \textit{longying\_team}\footnote[4]{LONGYINGZHIDA FINTECH}, and \textit{latim}\footnote[5]{Laboratoire de Traitement de l'Information Médicale}. Importantly, all the teams implemented and trained Unet-like architectures \cite{falk2019u} that did not see annotations from the second radiologist. Table \ref{ncct_metrics} summarizes the best achieved results per team w.r.t. the annotations of both radiologists in the test set. The \textit{icomaia} team obtained the best results for this task, localizing the most number of ischemic lesions (17 and 20 for R1 and R2, respectively) with a better DICE than other teams ($0.2$ for both radiologists). It should be noted that, \textit{xmu\_lsgroup} obtained the best HD for both radiologists (R1=$58.60$, R2=$60.33$), despite having a lower DICE. The achieved HD suggests a good lesion localization, being limited in the injured area estimation (low dice score). Additionally, the results on PREC and SENS suggest that models from \textit{longying\_team} and \textit{latim} are unable to detect lesions.

\begin{table}[h!]
\centering
\scalebox{0.64}{
\begin{tabular}{clccccc}
\toprule
\textbf{} & \textbf{Team} & \textbf{✓} & \textbf{DICE}   & \textbf{PREC}   & \textbf{SENS}   & \textbf{HD}        \\ \midrule
\multirow{4}{*}{R$_1$}  & \textit{icomaia}       & \textbf{17/36}      & \textbf{0.20 $\pm$ 0.26} & \textbf{0.18 $\pm$ 0.25} & \textbf{0.27 $\pm$ 0.32} & 67.68 $\pm$ 23.37  \\
                     & \textit{xmu\_lsgroup}   & 10/36      & 0.13 $\pm$ 0.32 & 0.14 $\pm$ 0.32 & 0.01 $\pm$ 0.04 & \textbf{58.60 $\pm$ 22.12}  \\
                     & \textit{longying\_team} & 7/36       & 0.12 $\pm$ 0.32 & 0.05 $\pm$ 0.19 & 0.00 $\pm$ 0.01 & 64.32 $\pm$ 22.99  \\
                     & \textit{latim}         & 6/36       & 0.11 $\pm$ 0.32 & 0.00 $\pm$ 0.01 & 0.00 $\pm$ 0.00 & 87.36 $\pm$ 31.85  \\ \midrule
\multirow{4}{*}{R$_2$}  & \textit{icomaia}       & \textbf{20/36}      & \textbf{0.20 $\pm$ 0.24} & \textbf{0.17 $\pm$ 0.24} & \textbf{0.29 $\pm$ 0.32} & 64.91 $\pm$ 24.78  \\
                     & \textit{xmu\_lsgroup}   & 10/36      & 0.10 $\pm$ 0.28 & 0.15 $\pm$ 0.34 & 0.01 $\pm$ 0.05 & \textbf{60.33 $\pm$ 23.38}  \\
                     & \textit{longying\_team} & 6/36       & 0.09 $\pm$ 0.28 & 0.04 $\pm$ 0.18 & 0.00 $\pm$ 0.01 & 78.01 $\pm$ 17.81  \\
                     & \textit{latim}         & 4/36       & 0.08 $\pm$ 0.28 & 0.00 $\pm$ 0.02 & 0.00 $\pm$ 0.00 & 101.30 $\pm$ 48.76 \\ 
\bottomrule
\end{tabular}
}
\caption{Summary of the NCCT segmentation metrics performance for the top four teams regarding the criteria both experts. The number in the check-mark indicate a correct classification of presence/abscense of the ischemic lesion.}
\label{ncct_metrics}
\end{table}

\begin{figure*}[h!]
    \centering
    \includegraphics[width=\textwidth]{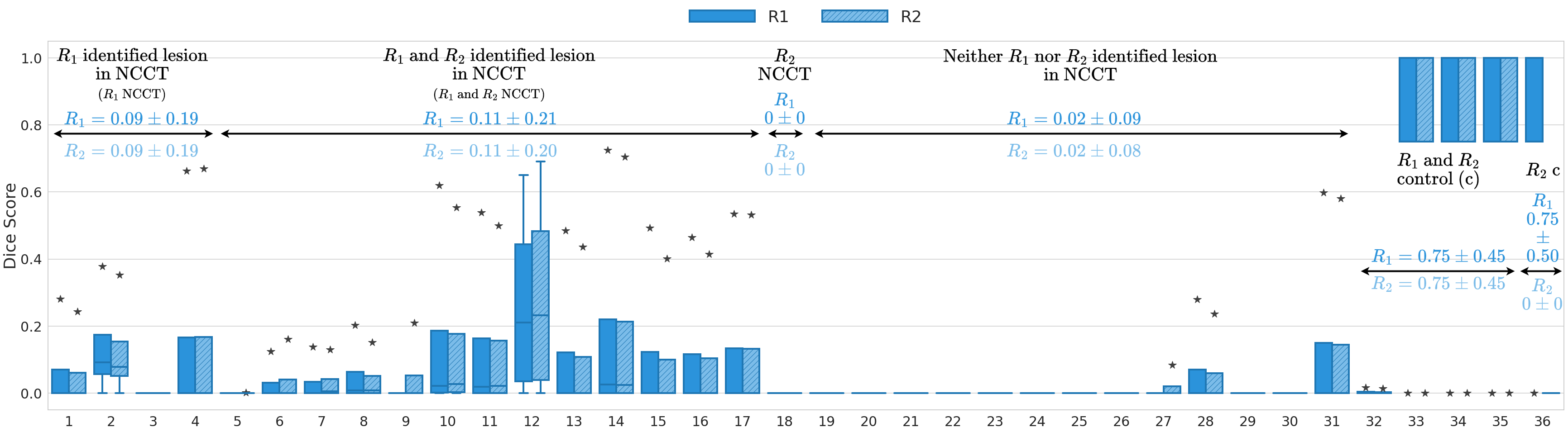}
    \caption{DICE scores for the participant strategies predictions over test NCCTs, against both expert annotations. The results are grouped based on the lesion identification task.}
    \label{ncct_boxplots}
\end{figure*}

Supported by the identification of lesions from NCCT, we categorized the test cases into groups of interest, shedding light on the challenges associated with detecting ischemic lesions from the CT modality. The groups are formed from samples where i) both experts agreed on lesion presence, ii) either R1 or iii) R2 identified a lesion, iv) neither R1 nor R2 found a lesion (excluding control), and also v) control. Hence, Figure \ref{ncct_boxplots} shows a summary of the achieved results for each patient, considering the annotations of both radiologists. Boxplots summarize the achieved dice scores for participants. As expected, both radiologists exhibit coherence on stroke identification, while the models are principally biased for control studies, and modelling of healthy regions. Regarding the stroke patients, as expected, the groups where the first radiologist was able to identify a lesion from NCCT obtained the best DICE scores (i=0.09, ii=0.11). Such results may be attributed to the fact that models were trained with the annotations of this radiologist. Interestingly, none of the solutions were able to segment the ischemic lesion from the sample with NCCT lesion identified by the second radiologist (third group). Additionally, from the fourth group, only in 23\% (n=3) of the samples the algorithms were able to detect stroke lesions, showing an average DICE of 0.02. 

Finally, as an optional and alternative task in the last week of the challenge, teams were invited to submit their strategies trained on ADC. This effort was made in order to compare the sensitivity offered by the two modalities available in the dataset. In this task, only three teams were able to submit an adjusted model for ADC studies. Table \ref{adc_metrics} shows that the ADC allows to identify with ease the presence/absence of ischemic stroke lesions (ADC=33, NCCT=17). The best result was achieved by \textit{xmu\_lsgroup} team, with a difference of 0.12 and 0.07 in DICE for the first and second radiologist over the second best strategy (\textit{longying\_team}), respectively. As observed, there is a remarked difference between the achieved segmentations on ADC and NCCT. This results are expected due to the greater sensitivity of ADC w.r.t. to acute findings, which in turn have been widely described in the literature.

\begin{table}[h!]
\centering
\scalebox{0.64}{
\begin{tabular}{clccccc}
\toprule
\textbf{}              & \textbf{Team}  & \textbf{✓} & \textbf{DICE}   & \textbf{PREC}   & \textbf{SENS}   & \textbf{HD}       \\ \midrule
\multirow{3}{*}{R$_1$} & \textit{xmu\_lsgroup}   & \textbf{33/36}      & \textbf{0.67 $\pm$ 0.29} & \textbf{0.65 $\pm$ 0.35} & \textbf{0.52 $\pm$ 0.33} & \textbf{23.67 $\pm$ 28.16} \\
                       & \textit{longying\_team} & 30/36      & 0.55 $\pm$ 0.35 & 0.62 $\pm$ 0.41 & 0.40 $\pm$ 0.34 & 25.77 $\pm$ 31.73 \\
                       & \textit{latim}          & 26/36      & 0.45 $\pm$ 0.36 & 0.56 $\pm$ 0.46 & 0.27 $\pm$ 0.28 & 25.84 $\pm$ 31.54 \\ \midrule
\multirow{3}{*}{R$_2$} & \textit{xmu\_lsgroup}   & \textbf{33/36}      & \textbf{0.58 $\pm$ 0.30} & \textbf{0.67 $\pm$ 0.32} & \textbf{0.45 $\pm$ 0.32} & 29.27 $\pm$ 32.05 \\
                       & \textit{longying\_team} & 30/36      & 0.51 $\pm$ 0.35 & 0.63 $\pm$ 0.39 & 0.38 $\pm$ 0.34 & \textbf{26.36 $\pm$ 31.53} \\
                       & \textit{latim}          & 20/36      & 0.41 $\pm$ 0.35 & 0.58 $\pm$ 0.45 & 0.26 $\pm$ 0.27 & 30.98 $\pm$ 29.24 \\
\bottomrule
\end{tabular}
}
\caption{Summary of the ADC segmentation metrics performance for the top four teams regarding the medical criteria R1 y R2.}
\label{adc_metrics}
\end{table}


\section{Discussion and concluding remarks}


This work summarized the results achieved on the APIS challenge, presented at ISBI2023 - the 20th IEEE International Symposium on Biomedical Imaging 2023. The APIS challenge motivated participant teams to design and implement models to segment stroke lesions over NCCT studies. A main motivation of this challenge was to bring paired information between NCCT and ADC studies, to learn and transfer lesion findings in both modalities. In the first stage, paired data was sent to the participants, together with stroke lesion annotations given by an expert radiologist. Then, the participants had to adjust the representation, which thereafter was validated only with NCCT inputs. Each model was validated regarding the annotations from two radiologists, and the best-achieved score was saved for each team. A total of 41 teams did an inscription to the challenge, but only four teams achieved to submit a solution in the proposed schedule. The winner team achieved a DICE score of 0.20 for both radiologists over NCCT studies, and a HD of 67.68 for the first radiologist and 64.91 for the second.

This challenge represented an additional effort to bring paired information of ADC and NCCT modalities to the scientific community. On one hand, the ADC has clear radiological findings that allow the localization of the lesion. In fact, the radiologist carried out the delineations over these studies. On the other hand, the NCCT studies remain as the principal image diagnosis that are firstly acquired from patients with stroke suspicion. From such open database is expected that scientific community find and desing novel mechanism to integrate modality information and produce computational tools to approximate lesion segmentations over NCCT studies, dealing with subjectivity reduction in shape lesion estimation. In the literature, other open databases and challenges dedicated to stroke lesion characterization have been proposed. For instance, the AISD dataset \cite{de_bruijne_symmetry-enhanced_2021} provides around 397 NCCT studies with associated delineations. Also, the ISLES2018 dataset provides 92 CTP studies, which is in turn involved in a challenge that has an additional 64 studies hidden for testing. Complementary, the  ISLES2018 provided parametric maps that highly support lesion estimation. Despite to these efforts to bring open data to model stroke lesions, we find a necessity to provide complementary information that allows us to better adjust representations to operate over NCCT studies. In fact, these studies only provide one delineation, which limit the validation of generalization capability of proposed alternatives.


From the results of the challenge, it was also evident the necessity to continue working in the design and modelling of computational representation to learn NCCT findings, which constitutes the first modality for stroke diagnosis. The best (DICE, HD) scores for R1 and R2 reference annotations were ($0.20, 58.60$) and ($0.20, 60.33$), respectively. This lower scores are comparable with results reported in the literature, making the segmentation limitations due to the poor contrast and hypoattenuation of the ischemic lesions. An additional evaluation was carried out over ADC studies, achieving DICE scores of 0.67 and 0.58 for the first and second radiologist, respectively. This remarkable difference is explained by the observation of the lesion in the modalities, however, MRI is not as widely used in clinical practice as CT, being the first line allowing triage.

Overall, the stroke segmentation over NCCT studies remains an open problem, which requires additional efforts to design new architectures and models to explore CT findings but also that take advantage of complementary paired information. Future works will include additional paired sequences in order to capture the variability of stroke patterns, where we hypothesize an improvement in the discrimination capabilities. In addition, exploring more preprocessing strategies allows to achieve a robust alignment between CT and MRI modalities and therefore, new NCCT to ADC translation strategies to produce synthetic images with an increased sensitivity to characterize the stroke lesions.

\section*{Acknowledgments}
The authors thank Ministry of science, technology and innovation of Colombia (MINCIENCIAS) for supporting this research work by the project "Mecanismos computacionales de aprendizaje profundo para soportar tareas de localización, segmentación y pronóstico de lesiones asociadas con accidentes cerebrovasculares isquémicos", with code 91934.

\bibliographystyle{IEEEbib}
\bibliography{refs-short}

\begin{thebibliography}{10}

\bibitem{feigin2022world}
Valery~L Feigin et~al.,
\newblock ``World stroke organization (wso): global stroke fact sheet 2022,''
\newblock {\em International Journal of Stroke}, vol. 17, no. 1, pp. 18--29,
  2022.

\bibitem{roth2020global}
Gregory~A Roth et~al.,
\newblock ``Global burden of cardiovascular diseases and risk factors,
  1990--2019: update from the gbd 2019 study,''
\newblock {\em Journal of the American College of Cardiology}, vol. 76, no. 25,
  pp. 2982--3021, 2020.

\bibitem{rennert_epidemiology_2019}
Robert~C Rennert et~al.,
\newblock ``Epidemiology, {Natural} {History}, and {Clinical} {Presentation} of
  {Large} {Vessel} {Ischemic} {Stroke},''
\newblock {\em Neurosurg.}, vol. 85, no. suppl\_1, pp. S4--S8, July 2019.

\bibitem{powers2019guidelines}
William~J Powers et~al.,
\newblock ``Guidelines for the early management of patients with acute ischemic
  stroke: 2019 update to the 2018 guidelines for the early management of acute
  ischemic stroke: a guideline for healthcare professionals from the american
  heart association/american stroke association,''
\newblock {\em Stroke}, vol. 50, no. 12, pp. e344--e418, 2019.

\bibitem{bevers2018apparent}
Matthew~B Bevers et~al.,
\newblock ``Apparent diffusion coefficient signal intensity ratio predicts the
  effect of revascularization on ischemic cerebral edema,''
\newblock {\em Cerebrovascular Diseases}, vol. 45, no. 3-4, pp. 93--100, 2018.

\bibitem{Martel1999}
Anne~L. Martel et~al.,
\newblock ``{Measurement of infarct volume in stroke patients using adaptive
  segmentation of diffusion weighted MR images},''
\newblock in {\em Lecture Notes in Computer Science (including subseries
  Lecture Notes in Artificial Intelligence and Lecture Notes in
  Bioinformatics)}, 1999.

\bibitem{rana2003apparent}
Arnab~K Rana et~al.,
\newblock ``Apparent diffusion coefficient (adc) measurements may be more
  reliable and reproducible than lesion volume on diffusion-weighted images
  from patients with acute ischaemic stroke--implications for study design,''
\newblock {\em Magnetic resonance imaging}, vol. 21, no. 6, pp. 617--624, 2003.

\bibitem{maier_isles_2017}
Oskar Maier et~al.,
\newblock ``Isles 2015-a public evaluation benchmark for ischemic stroke lesion
  segmentation from multispectral mri,''
\newblock {\em Medical image analysis}, vol. 35, pp. 250--269, 2017.

\bibitem{winzeck2018isles}
Stefan Winzeck et~al.,
\newblock ``Isles 2016 and 2017-benchmarking ischemic stroke lesion outcome
  prediction based on multispectral mri,''
\newblock {\em Frontiers in neurology}, vol. 9, pp. 679, 2018.

\bibitem{hakim2021predicting}
Arsany Hakim et~al.,
\newblock ``Predicting infarct core from computed tomography perfusion in acute
  ischemia with machine learning: Lessons from the isles challenge,''
\newblock {\em Stroke}, vol. 52, no. 7, pp. 2328--2337, 2021.

\bibitem{petzsche2022isles}
Moritz~R Hernandez~Petzsche et~al.,
\newblock ``Isles 2022: A multi-center magnetic resonance imaging stroke lesion
  segmentation dataset,''
\newblock {\em Scientific data}, vol. 9, no. 1, pp. 762, 2022.

\bibitem{liew2018large}
Sook-Lei Liew et~al.,
\newblock ``A large, open source dataset of stroke anatomical brain images and
  manual lesion segmentations,''
\newblock {\em Scientific data}, vol. 5, no. 1, pp. 1--11, 2018.

\bibitem{liew2022large}
Sook-Lei Liew et~al.,
\newblock ``A large, curated, open-source stroke neuroimaging dataset to
  improve lesion segmentation algorithms,''
\newblock {\em Scientific data}, vol. 9, no. 1, pp. 320, 2022.

\bibitem{de_bruijne_symmetry-enhanced_2021}
Kongming Liang et~al.,
\newblock ``Symmetry-enhanced attention network for acute ischemic infarct
  segmentation with non-contrast ct images,''
\newblock in {\em Medical Image Computing and Computer Assisted
  Intervention--MICCAI 2021: 24th International Conference, Strasbourg, France,
  September 27--October 1, 2021, Proceedings, Part VII 24}. Springer, 2021, pp.
  432--441.

\bibitem{Tureckova2019}
Alzbeta Tureckova et~al.,
\newblock ``{ISLES challenge: U-shaped convolution neural network with dilated
  convolution for 3D stroke lesion segmentation},''
\newblock {\em Lecture Notes in Computer Science}, vol. 11383 LNCS, pp.
  319--327, 2019.

\bibitem{Kuang2021}
Hulin Kuang et~al.,
\newblock ``{EIS-Net: Segmenting early infarct and scoring ASPECTS
  simultaneously on non-contrast CT of patients with acute ischemic stroke},''
\newblock {\em Medical Image Analysis}, vol. 70, no. February, pp. 101984,
  2021.

\bibitem{Wang2020}
Guotai Wang et~al.,
\newblock ``{Automatic ischemic stroke lesion segmentation from computed
  tomography perfusion images by image synthesis and attention-based deep
  neural networks},''
\newblock {\em Medical Image Analysis}, vol. 65, pp. 101787, 2020.

\bibitem{gutierrez_lesion-preserving_2023}
Alejandro Gutierrez et~al.,
\newblock ``Lesion-preserving unpaired image-to-image translation between {MRI}
  and {CT} from ischemic stroke patients,''
\newblock {\em Int J CARS}, vol. 18, no. 5, pp. 827--836, Jan. 2023.

\bibitem{kuang2019automated}
Hulin Kuang et~al.,
\newblock ``Automated infarct segmentation from follow-up non-contrast ct scans
  in patients with acute ischemic stroke using dense multi-path contextual
  generative adversarial network,''
\newblock in {\em Medical Image Computing and Computer Assisted
  Intervention--MICCAI 2019: 22nd International Conference, Shenzhen, China,
  October 13--17, 2019, Proceedings, Part III 22}. Springer, 2019, pp.
  856--863.

\bibitem{wang2016deep}
Yanran Wang et~al.,
\newblock ``A deep symmetry convnet for stroke lesion segmentation,''
\newblock in {\em 2016 IEEE International Conference on Image Processing
  (ICIP)}. IEEE, 2016, pp. 111--115.

\bibitem{dekeyzer2017distinction}
Sven Dekeyzer et~al.,
\newblock ``Distinction between contrast staining and hemorrhage after
  endovascular stroke treatment: one ct is not enough,''
\newblock {\em Journal of NeuroInterventional Surgery}, vol. 9, no. 4, pp.
  394--398, 2017.

\bibitem{wang_asymmetry_2022}
Haomiao Ni et~al.,
\newblock ``Asymmetry {Disentanglement} {Network} for {Interpretable} {Acute}
  {Ischemic} {Stroke} {Infarct} {Segmentation} in {Non}-contrast {CT}
  {Scans},''
\newblock in {\em Medical {Image} {Computing} and {Computer} {Assisted}
  {Intervention} – {MICCAI} 2022}, vol. 13438, pp. 416--426. Springer Nature
  Switzerland, Cham, 2022.

\bibitem{xu_combining_2023}
Zhixiang Xu and Changsong Ding,
\newblock ``Combining convolutional attention mechanism and residual deformable
  {Transformer} for infarct segmentation from {CT} scans of acute ischemic
  stroke patients,''
\newblock {\em Front. Neurol.}, vol. 14, pp. 1178637, July 2023.

\bibitem{Dolz2019}
Jose Dolz et~al.,
\newblock ``{Dense multi-path u-net for ischemic stroke lesion segmentation in
  multiple image modalities},''
\newblock {\em Lecture Notes in Computer Science (including subseries Lecture
  Notes in Artificial Intelligence and Lecture Notes in Bioinformatics)}, vol.
  11383 LNCS, pp. 271--282, 2019.

\bibitem{Clerigues2019}
Albert Cl{\`{e}}rigues et~al.,
\newblock ``{Acute ischemic stroke lesion core segmentation in CT perfusion
  images using fully convolutional neural networks},''
\newblock {\em Computers in Biology and Medicine}, vol. 115, pp. 103487, 2019.

\bibitem{Liu2019}
Pengbo Liu,
\newblock ``{Stroke lesion segmentation with 2D novel CNN pipeline and novel
  loss function},''
\newblock {\em Lecture Notes in Computer Science (including subseries Lecture
  Notes in Artificial Intelligence and Lecture Notes in Bioinformatics)}, vol.
  11383 LNCS, pp. 253--262, 2019.

\bibitem{allmendinger2012imaging}
Andrew~Mark Allmendinger et~al.,
\newblock ``Imaging of stroke: Part 1, perfusion ct -\- overview of imaging
  technique, interpretation pearls, and common pitfalls,''
\newblock {\em American Journal of Roentgenology}, vol. 198, no. 1, pp. 52--62,
  2012.

\bibitem{gauriau_head_2023}
Romane Gauriau et~al.,
\newblock ``Head ct deep learning model is highly accurate for early infarct
  estimation,''
\newblock {\em Scientific Reports}, vol. 13, no. 1, pp. 189, 2023.

\bibitem{li2016first}
Xiangrui Li et~al.,
\newblock ``The first step for neuroimaging data analysis: Dicom to nifti
  conversion,''
\newblock {\em Journal of neuroscience methods}, vol. 264, pp. 47--56, 2016.

\bibitem{falk2019u}
Thorsten Falk et~al.,
\newblock ``U-net: deep learning for cell counting, detection, and
  morphometry,''
\newblock {\em Nature methods}, vol. 16, no. 1, pp. 67--70, 2019.

\end{thebibliography}

\end{document}